# Broadened phonon-assisted wide-band radiation and subsequent low-threshold self-absorption coherent modulation in the high-entropy glass system doped with Nd$^{3+}$ ions

Linde Zhang[1], Jingyuan Zhang[1], Xiang Wang[2], Meng Tao[3], Gangtao Dai[4], Jing Wu[3], Zhangwang Miao[1], Shifei Han[1], Haijuan Yu[1], Xuechun Lin[1, ✉]

[1]Laboratory of All-solid-state Light Sources, Beijing Engineering Research Center, Institute of Semiconductors, Chinese Academy of Sciences, Beijing, China
[2]Synlumin Conuninex (Shanghai) Enterprise Development Co., Ltd., Shanghai, China
[3]Time-wave-space Optical Technology(Xiaogan) Co., Ltd., Xiaogan, Hubei, China
[4]High-dimensional Plasma Sources Technology(Xiaogan) Co., Ltd., Xiaogan, Hubei, China
✉e-mail: xclin@semi.ac.cn

## Abstract

For crystalline materials with long-range orders, the phonon modes involved in the phonon-assisted radiation process are generally involving one or several phonons with specific vibration frequencies [1-4]. In some glassy material, the phonon modes broaden in a short range [5-7]. However, the locally distinct chemical environments or mass disorder in high-entropy systems can induce an anharmonic phonon-phonon coupling and composition disorder, which leads to significant phonon broadening [8,9]. The terminology of high-entropy comes from the high configuration entropy larger than 1.5R (R is the gas constant), which results from randomly distributed multiple nearly equal components in a crystal lattice [10,11]. Inspired by the high-entropy strategy, we deployed a high-entropy glass system (HEGS) doped with neodymium ions, which exhibits a complex structure with tetrahedral voids filled by different ions, including Li$^+$, Zn$^{2+}$, Si$^{4+}$, P$^{5+}$, S$^{6+}$, etc. Phonon's spectral broadening up to thousands of wavenumbers in the HEGS allows strong wide-band absorption in both the near-infrared and mid-infrared ranges and assists the system's radiation, i.e., broadened phonon-assisted wide-band radiation (BPAWR). The subsequent low-threshold self-absorption coherence modulation (SACM) was also observed in the HEGS, modulated by changing excitation wavelengths, sample size, and doping concentrations. The time delay of the BPAWR signal is up to 1.66 ns relative to the zero-delay signal, while the time delay of the Raman process is typically in the order of fs to ps, rarely up to 10 ps [12-15]. The BPAWR-SACM can be applied to realize signal amplification of the centered non-absorption band when dual-wavelength lasers pump the HEGS sample, and signal amplification can be up to 26.02 dB. The spectral characteristics of the BPAWR and the dynamics of the energy distribution of the excited species are investigated in detail. It can be expected that such a novel radiative de-excitation process helps to achieve tunability of optical fiber laser and paves the pathway for numerous new applications in supercontinuum, laser amplification, and optical communication.

The vibrational modes of solid matrix often directly involved in the de-excitation process of the excited center, i.e., phonon-assisted radiation if depicted by a quasi-particle picture [1]. When the phonon energy of the matrix is high, and the Huang-Rhys factor between the matrix and the excited center is large, the typical non-radiative transition process, such as the multi-phonon emission, can even dominate the de-excitation process [2,16]. When the excitation energy is sufficiently high, it can pump the excited center from the ground state to one of the excited states with a certain energy level. The force constant of the vibration modes in the excited state is lower than that in the ground state. Such a situation would enhance the phonon-electron coupling, causing more phonons to contribute to the electronic transition between the two states and improving the transition polarizability tensor. It is the well-known resonant Raman scattering [17-20]. In the radiative de-excitation process of excitons, the phonons can change the band structure locally due to the deformation of the atomic lattice (deformation-potential scattering), and it is so-called phonon side band emission [21]. So far, phonon scattering was mainly used to explore the thermal and transport characteristics of high entropy materials [9,22-24].

We applied the high-entropy strategy in an alternative glass phase system to realize a class of $Nd^{3+}$ ions doped high-entropy glass system (HEGS). It exhibits a complex structure with tetrahedral voids filled by different ions, including $Li^+$, $Zn^{2+}$, $Si^{4+}$, $P^{5+}$, $S^{6+}$, etc. Hence, the configuration entropy of the system significantly increases, even greater than the melting entropy of the system itself [25]. The digital image (Fig. 1a) shows the HEGS rods with varying concentrations. The intensity-interplanar relation (Fig. 1b) proves that the HEGS has a weakly ordered structure. The inset inside shows that no particulates precipitation is present in the HEGS, and the overall contrast remains consistent. An endothermic peak in the differential scanning calorimetry (DSC) result (Fig. 1c) can be explained by enthalpy relaxation [26,27]. Near the vitrification temperature, the relatively ordered structure in the HEGS disintegrates, and the corresponding free volume expands, which causes enthalpy relaxation and a non-reversing endothermic peak. This also corresponds to a non-convergent order-disorder phase transition [28]. Above 400 $cm^{-1}$, the absorption capacity of the HEGS is up to 100%, ending at 3600 $cm^{-1}$ (Fig. 1d). The infrared absorption fades to zero at 6500 $cm^{-1}$, the infrared absorption side band, which is much higher than that of other glass systems [5-7]. The wide-band infrared absorption results from significantly broadened phonon modes in the HEGS, which makes infrared with different frequencies excite corresponding phonon modes. Phonon broadening in the HEGS also leads to optical phonon modes with frequencies much higher than that in conventional solid matrixes [29,30].

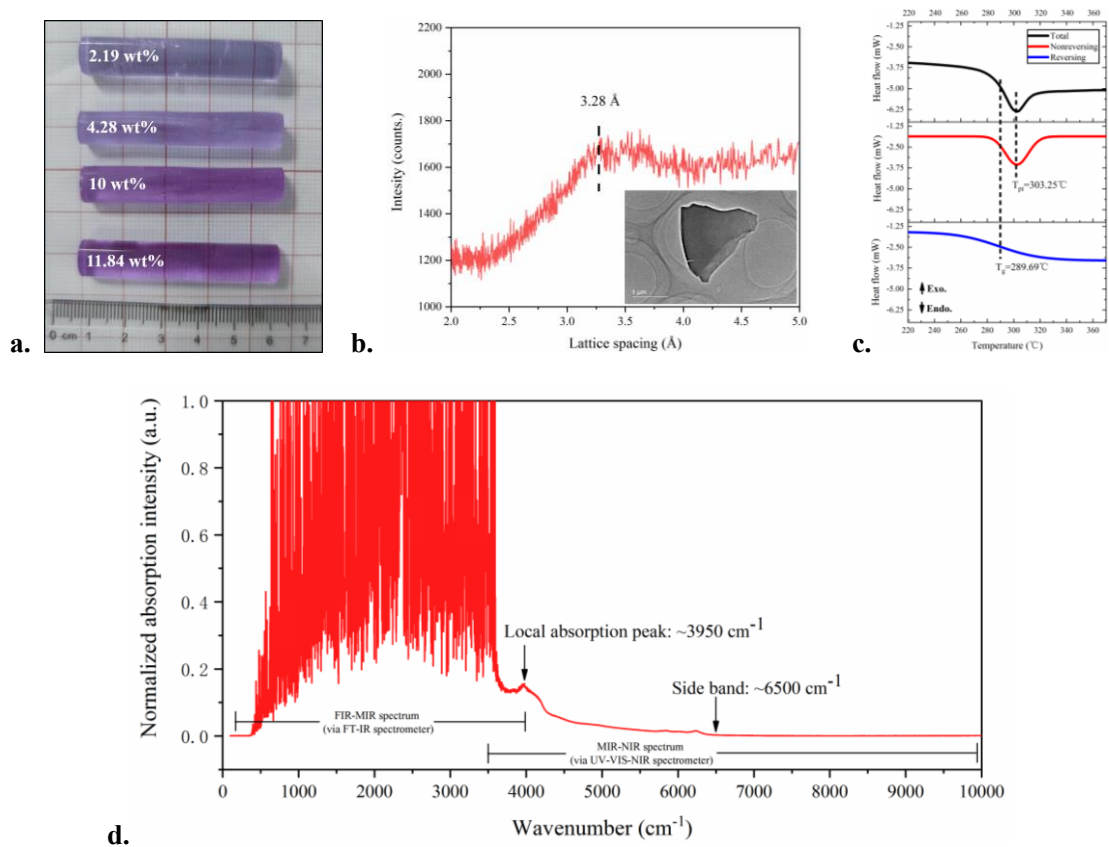

**Figure 1 | Structure, thermodynamics and spectral properties of HEGS. a,** Photo of HEGS samples with varying concentrations. **b,** The intensity-interplanar relationship of the HEGS system, derived from the original intensity-diffraction angle relationship from X-ray diffraction measurement. The inset is the TEM micrograph of the

HEGS sample. **c,** DSC result of one HEGS sample. When the sample was heated to around 300°C, a non-reversing endothermic peak is present, accompanying the reversing vitrification process. Isolation of the two processes was accomplished by peak fitting. **d,** The extended absorption spectrum of HEGS from far-infrared to near-infrared. FIR-MIR absorption spectrum was obtained by converting the FT-IR transmission spectrum to absorption spectrum. The MIR-NIR absorption spectrum was obtained by UV-VIS-NIR measurement. Normalizing the two absorption spectra and combining them to get the extended absorption spectrum.

The high-frequency optical phonon modes play an important role in the phonon-assisted radiation process in the visible light range. Apart from residual excitation light signals, the emission spectrum (upper part in Fig. 2a) also includes some new frequency components that show significant red-shift or blue-shift relative to the excitation frequency and can be directly observed in the inset. These new frequency components have no relation with the fluorescence of $Nd^{3+}$ ions since they cannot be attributed to the emission spectrum of $Nd^{3+}$ transition between any two energy levels [31,32]. Moreover, the red-shift and blue-shift ranges are up to about 10 000 $cm^{-1}$ and about 7 000 $cm^{-1}$, respectively. The asymmetric shift is different from conventional Raman shift [33,34]. It can also be found that there is a certain correlation between the emission spectrum and the absorption spectrum (lower part in Fig. 2a). The peaks indicated by black arrows represent the excitation process of $Nd^{3+}$ ions transition from the ground state to the excited state. Extremely low emission intensity is detected, corresponding to these absorption peaks. In contrast, the peaks indicated by red arrows correspond to no absorption or weak absorption in the absorption spectrum. It is supposed that the newly generated red-shifted and blue-shifted components compose quasi-continuous wide-band emission peaks, which corresponds to a broadening phonon-assisted wide-band radiation (BPAWR) process due to de-excitation of $Nd^{3+}$ ions in the excited states. However, due to the self-absorption of some frequency components by $Nd^{3+}$ ions, the wide-band emission peaks are fragmented into a series of different frequency components, exhibiting discontinuous and asymmetric features. Therefore, the newly generated frequency components result from the combined effect of BPAWR and the subsequent self-absorption modulation (SAM).

Experimentally, we measured the emission spectra at different angles relative to the forward direction of the excitation lasers for spatial coherence study (Fig. 2b, Extended data Fig. 1a,b). The results show that only when the measurement is conducted at 0° can the red-shifted and blue-shifted BPAWR-SAM signal be observed. If the measurement angle turns 45°, a weak scattering excitation signal and an emission peak at 11199 $cm^{-1}$ are present. This emission peak can be attributed to the corresponding fluorescence emission of $Nd^{3+}$ ions de-excited from $^4F_{3/2}$ to $^4I_{9/2}$. In detail, the $Nd^{3+}$ ions are excited to energy level $^4F_{5/2}$ from $^4I_{9/2}$, de-excited to energy level $^4F_{3/2}$ by phonon emission, and return to ground state $^4I_{9/2}$ by emitting light. At the measurement angles of 135° and 180°, the fluorescence emission peak intensity from $^4F_{3/2}$ to $^4I_{9/2}$ is extremely weak. These results prove that the BPAWR-SAM process has strong spatial coherence. Since the BPAWR process alone has no spatial coherence, the observed spatial coherence should be attributed to SAM [35]. Therefore, the SAM process can be named self-absorption coherence modulation (SACM), and the overall process is referred to as BPAWR-SACM.

The time delay study between the excitation laser and the generated radiation proved that the BPAWR process differs from the conventional radiation process (Fig. 2c). The corresponding optical experiment setup and obtained BPAWR-SACM spectra were presented (Supplementary Fig. 1,2). With the triggering signal from the pump as a reference, the time delay corresponding to the zero-delay signal and the BPAWR-SACM signal is -0.43 ns and 1.23 ns, respectively. Hence, the time delay of the BPAWR-SACM signal is 1.66 ns relative to the zero-delay signal. As for the conventional Raman and stimulated Raman process, the time delay is in the order of fs to ps, rarely up to 10 ps [12-15].

We also studied the conversion efficiency of the BPAWR-SACM process by measuring the ratio between the output power and the pump laser power (Fig. 2d), and the emission spectra were also acquired (Extended data Fig. 2a,b). With the 874.9 nm laser pumping at 8.5 mW, the corresponding BPAWR-SACM power is relatively weak (only 0.05 mW), but the conversion efficiency is as high as 0.58%. By increasing laser power, the BPAWR-SACM power increases, reaching 0.52 mW when the pump power is 770 mW. However, the conversion efficiency decreases dramatically to 0.07%. With the 804.2 nm laser pumping at 13 mW, the corresponding BPAWR-SACM power is lower than the instrument's detection limit. It increased to 0.205 mW when the pump power was 700 mW. The conversion efficiency increased with increasing laser power, and the maximum value was 0.03%. The results support that the BPAWR-SACM process has an extremely low threshold, and laser power in the order of ten milliwatts is enough to excite the process. In contrast, the conventional nonlinear process needs extremely high light intensity for signal generation [36]. The varied conversion efficiency at different excitation wavelengths may come from the different absorption intensities.

We proposed a mechanism for the peculiar BPAWR-SACM process to facilitate the interpretation of the process (Fig. 2e). When an excitation laser excites the excited centers in the HEGS medium to one excited state having a real energy level, it greatly enhances the phonon-electron interaction between the excited centers and the medium. Many excited centers are de-excited by phonon emission and successively emit photons with varying frequencies. Since the phonons can cover a broad spectrum up to thousands of wavenumbers, the corresponding emission spectrum of the phonon-assisted radiation process also shows broadening by thousands of wavenumbers. Such a process is the so-called BPAWR described above, which contains several absorption spectra between different electronic and surrounding vibration-rotation excited states. When the BPAWR process occurs in the excited center, the ground state adjacent to the excited center can directly absorb a specific wavelength and transit to the corresponding excited state, leading to the photons' self-absorption process. The excited state returns to the ground state by a multi-phonon assisted process or emitting a photon at a shifted frequency with phonons' assistance. Regardless of the de-excitation route, the photons' self-absorption process will lead to a loss of the BPAWR emission until it exhibits a discontinuous spectrum. Meanwhile, the phonons produced by the de-excitation process will be absorbed by the following excited states, which is the phonons' self-absorption process. The wide-band radiation is permitted to propagate only in the non-absorption band of the medium, which consists of the blue-shifted band produced by the phonons' self-absorption and the red-shifted band produced by the phonons' emission. Both of them cause stimulated radiation from other excited states when propagating within the medium, thus amplifying the signal intensity of the residual band in the BPAWR process and imparting coherent properties to the corresponding band. This process is so-called SACM. The frequency components are changed accompanying the radiation propagation in the SACM process, which causes mode competition in the radiation process [37]. The SACM process induced by selecting absorption modes causes the phonon modes involved in the BPAWR process to depend on the surrounding density of allowable excited states. That is why the BPAWR and SACM processes are inseparable.

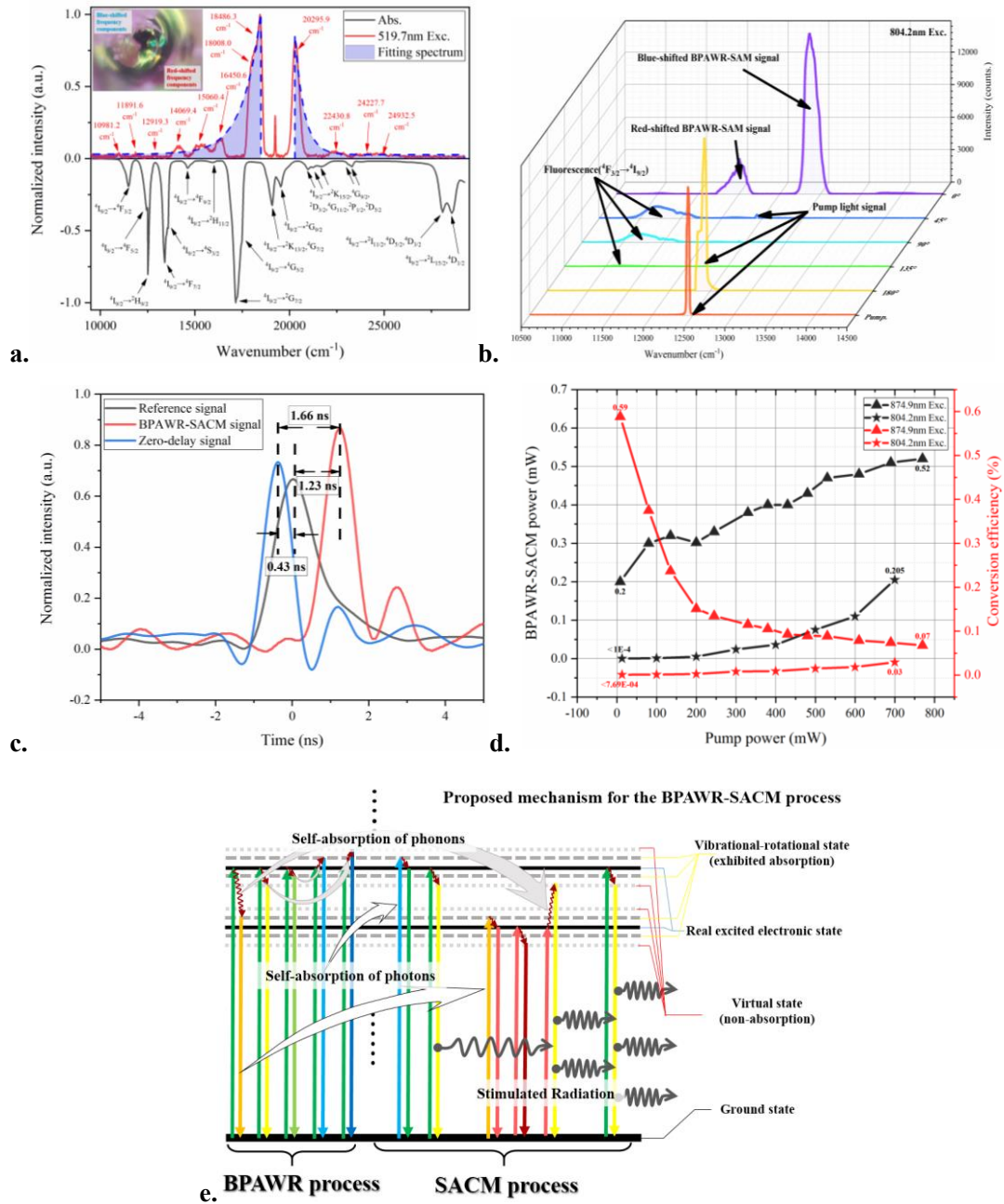

**Figure 2 | Photo-physical properties of BPAWR-SACM process. a,** Normalized emission spectrum (upper part) and normalized absorption spectrum (lower part) in the visible light range. The corresponding transition modes of $Nd^{3+}$ ions are marked. The fitting spectrum represents the hypothesized emission spectrum with non-absorption. The inset is a digital radiation image of the HEGS sample captured from the opposite side of the incident laser. **b,** Emission spectra at different measurement angles relative to the forward direction of the excitation light. The corresponding excited energy levels are marked in the figure. **c,** Time-delay measurement between the pump and the generated frequency-shifted radiation. Zero-delay corresponds to the arrival time of the pump pulse. **d,** BPAWR-SACM signal power and conversion efficiency versus pump power of laser excitation at 874.9 nm (denoted as 874.9 nm Exc.) and 804.2 nm (denoted as 804.2 nm Exc.). **e,** Suggested mechanism for the BPAWR-SACM process.

The proposed mechanism of BPAWR-SACM in Fig. 2e suggests that the excitation wavelengths that can induce the BPAWR-SACM process correspond to the strong absorption in the absorption spectrum of $Nd^{3+}$ ions transition in each energy level. The weak absorption in the absorption spectrum cannot induce the BPAWR-SACM process, even BPAWR alone. We studied the influence of excitation wavelengths in the BPAWR-SACM process by measuring the emission spectra (upper part in Fig. 3a) and extracting the spectral shift of the BPAWR-SACM signals relative to the excitation light (Supplementary Fig. 3). When the sample is excited by lasers at 441.6 nm, 531.2 nm, and 636.0 nm, only the residual excitation signals are observed. In contrast, with other excitation sources, the residual excitation signals are relatively weak or unobservable, but the red-shifted or blue-shifted signals are present. Laser at 515.2 nm, 804.2 nm, and 874.9 nm can excite $Nd^{3+}$ from $^4I_{9/2}$ in the ground state to $^4G_{9/2}$, $^4F_{5/2}$, and $^4F_{3/2}$ in the excited state, respectively. Further statistics to the data in Fig. 3a show that the area ratio of red-shifted to blue-shifted BPAWR-SACM signal at different excitation wavelengths is positively correlated with the absorption peak area-ratio between adjacent higher excited state and lower excited state (Fig. 3b). The above results can be attributed to the fact that the surrounding energy level density determines the area and intensity of one emission peak, as suggested in Fig. 2e. These results support that by inducing BPAWR-SACM with lasers of different wavelengths, modulating radiation frequency within thousands of wavenumbers is possible.

Beer-Lambert law describes that the absorption intensity is proportional to the optical path length and the doping concentration of the medium. It inspires us to tune the SACM process by varying the sample's length, and the corresponding emission spectra were shown in Fig. 3c. For a 10 mm sample, the red-shifted or blue-shifted BPAWR-SACM signals are not observed, but only a strong signal of the residual excitation light. When the glass length is extended to 60 mm, a red-shift at 770 $cm^{-1}$ and blue-shift at 1074 $cm^{-1}$ appear. Now the intensity of red-shifted or blue-shifted BPAWR-SACM signals reaches the maximum and starts to decrease if the sample becomes longer. For a 70 mm sample, the excitation signal disappears, and the BPAWR-SACM signals continue attenuating. The results suggest that when an excitation laser enters the HEGS medium, it is absorbed and attenuated continuously during propagation. Such an attenuation process accompanies the generation of the BPAWR-SACM before the excitation is completely consumed. At the same time, the newly generated red-shifted and blue-shifted signals continue to propagate forward until the BPAWR-SACM signals leave the HEGS or are completely absorbed by the medium. The optical path difference between the excitation laser and the BPAWR-SACM signal is around 60 mm. The light attenuation process supports that the BPAWR-SACM process in the HEGS is not formed instantaneously but is accumulated and gradually enhanced during the propagation. Therefore, the intensity ratio of emission peaks can be regulated by changing the optical path's length.

We also investigated the effect of the doping concentration of $Nd^{3+}$ ions on the BPAWR-SACM process (Fig. 3d). When the doping concentration is 2.19 wt.%, the excitation light intensity is strong, while the blue-shifted BPAWR-SACM signal is too weak to be distinguished from the signal noise. The strongest red-shifted BPAWR-SACM signal is located at 2666 $cm^{-1}$, much weaker than that of the excitation light. With increasing doping concentration, the intensity of BPAWR-SACM signals is enhanced. Some relatively weak red-shifted or blue-shifted BPAWR-SACM peaks can also be observed in the emission spectra. When the doping concentration reaches 11.84 wt.%, the intensity of red-shifted or blue-shifted BPAWR-SACM signals is greatly enhanced and becomes much higher than that of the residual excitation light. The above data support that increasing doping concentration enhances the absorption intensity of excitation light of the system.

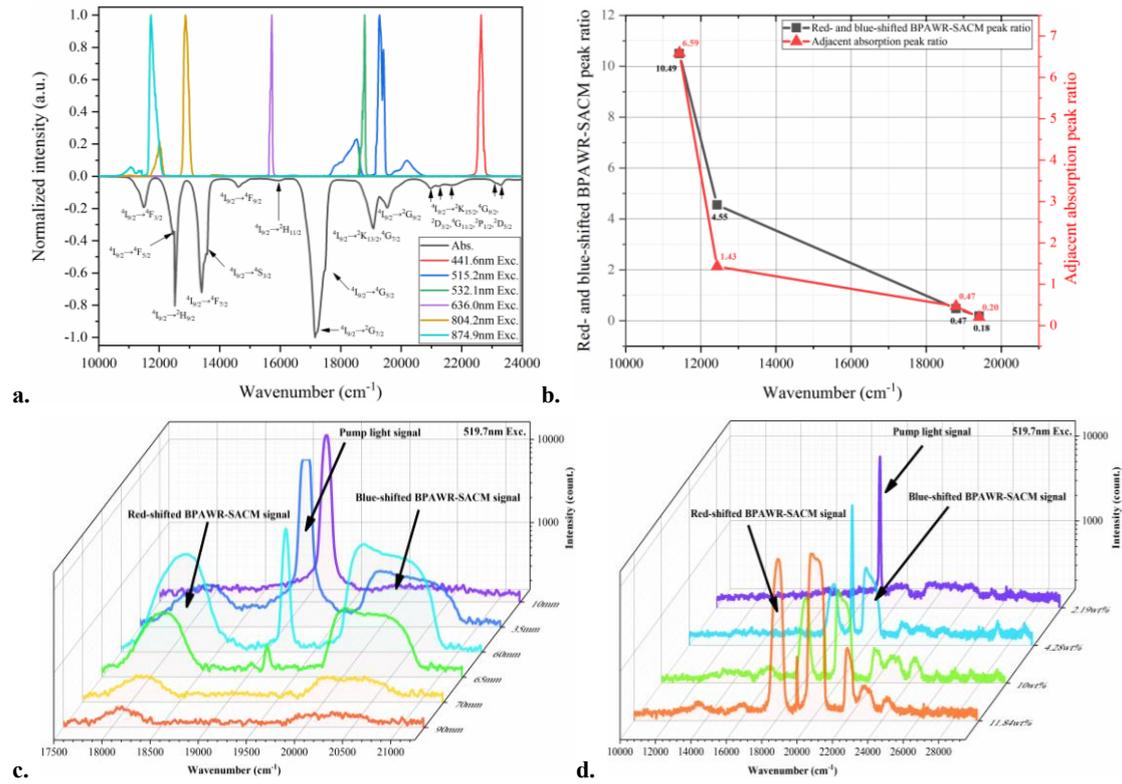

**Figure 3 | Modulation results of BPAWR-SACM process based on absorption characteristics. a,** Normalized emission spectra at different excitation wavelengths (upper part) and normalized absorption spectrum in the visible light range (lower part). Corresponding excited energy levels are marked in the figure. **b,** Statistical results of the red-shifted and blue-shifted BPAWR-SACM peak area ratio versus the adjacent absorption peak area ratio. All relevant excited states within phonon frequency of 3000 cm$^{-1}$ were considered in the statistics. **c,** Emission spectra measured by 519.7 nm excitation at samples with different lengths. **d,** Emission spectra of samples with different doping concentrations excited by 519.7 nm laser. The intensities in (**c**) and (**d**) were presented in logarithm to facilitate the observation of relatively weak emission peaks.

In the conventional stimulated Raman [38,39], a combination of specific wavelengths is needed for signal amplification. As long as certain absorption intensity is present before and/or after the non-absorption band, the energy of the dual- or even multi-wavelength pump lasers can be transferred to the non-absorption spectrum for signal amplification [40]. We demonstrated one kind of signal amplification in the centered non-absorption band by a dual-wavelength pump in this work. The work was done by keeping the power of one laser at 100 mW and varying the power of another laser alone (Fig. 4a, Extended data Fig. 3) or controlling the dual-wavelength pump power at the same level (Fig. 4b). As shown in Fig. 4a, when the 874.9 nm laser is off, the red-shifted 804.2 nm excited BPAWR-SACM signal starts at 12130.2 cm$^{-1}$ (i.e., 824.4 nm), reaches a peak at 12004.9 cm$^{-1}$, and ends at 11701.1 cm$^{-1}$. This signal, viewed as the target signal, is significantly weaker than the blue-shifted BPAWR-SACM signal. With a pump power of 500 mW of 874.9 nm laser, the peak of the target signal is found at 11757.9 cm$^{-1}$, showing a red-shift. Correspondingly, the target signal ends at 11589.3 cm$^{-1}$ (i.e., 862.9 nm). Compared with the 804.2 nm laser, the laser at 874.9 nm effectively amplifies the target signal.

As shown in Fig. 4b, at a power of 100 mW, the target signal starts at 12130.2 cm$^{-1}$, reaches a peak at 11696.9 cm$^{-1}$, and ends at 11589.3 cm$^{-1}$. As the dual-wavelength pump power is increased simultaneously, the target signal region remains unchanged, but the peak position is slightly blue-shifted. When the power reaches 300 mW, the peak is found at 11757.9 cm$^{-1}$. Taking the red-shifted component from 874.9 nm excitation and the blue-shifted component from 804.2 nm excitation of the BPAWR-SACM as a reference, one can find that the target signal region can be significantly enhanced. These results suggest that when two lasers having similar frequencies are incident on the HEGS sample, most energies are transferred to the centered non-absorption band and significantly amplify the signal in the range of 824.4 nm to 862.9 nm. For the dual-wavelength pump, the blue-shifted component at 874.9 nm excited BPAWR-SCAM signal between 824.4 nm and 862.9 nm can serve as a seed light [41] and greatly amplify the red-shifted 804.2 nm excited BPAWR-SACM signal. As a result, the signal amplification occurred at the centered non-absorption band, which included amplification from red-shifted 804.2 nm excited BPAWR-SACM signal and the blue-shifted 874.9 nm excited BPAWR-SACM signal. Increasing the pump powers of the two wavelengths synchronously can also minimize the displacement of the central peak position, thereby avoiding inconsistent gain effects.

Based on the BPAWR-SACM signals (Fig. 4a,b, Extended data Fig. 3), the gain of one laser relative to another laser and the average gain of the dual-wavelength pump were calculated (Fig. 4c). When the red-shifted 804.2 nm Exc BPAWR-SACM signal served as a target signal, the gain from the amplification of the blue-shifted 874.9 nm Exc BPAWR-SACM signal increased with increasing power, and the maximum reached 26.02 dB. A similar trend was found when the blue-shifted 874.9 nm Exc BPAWR-SACM signal is served as a target signal. But the gain from the amplification of the red-shifted 804.2 nm Exc BPAWR-SACM signal only had a maximum value of 1.14 dB. When the dual-wavelength pump power was increased synchronously, overall, the average gain increased with increasing power, and the maximum gain reached 12.39 dB. The dual-wavelength pump can significantly amplify the centered non-absorption band between the dual wavelengths. The amplification efficiency of the blue-shifted BPAWR-SACM signal is higher than that of the red-shifted BPAWR-SACM signal. Such a phenomenon is mainly attributed to the fact that the absorption of the 804.2 nm excitation is much stronger than the excitation at 874.9 nm. The corresponding conversion efficiency at 804.2 nm excitation is also lower (Fig. 2d). Resultantly, the gain is higher in the case of 874.9 nm excitation.

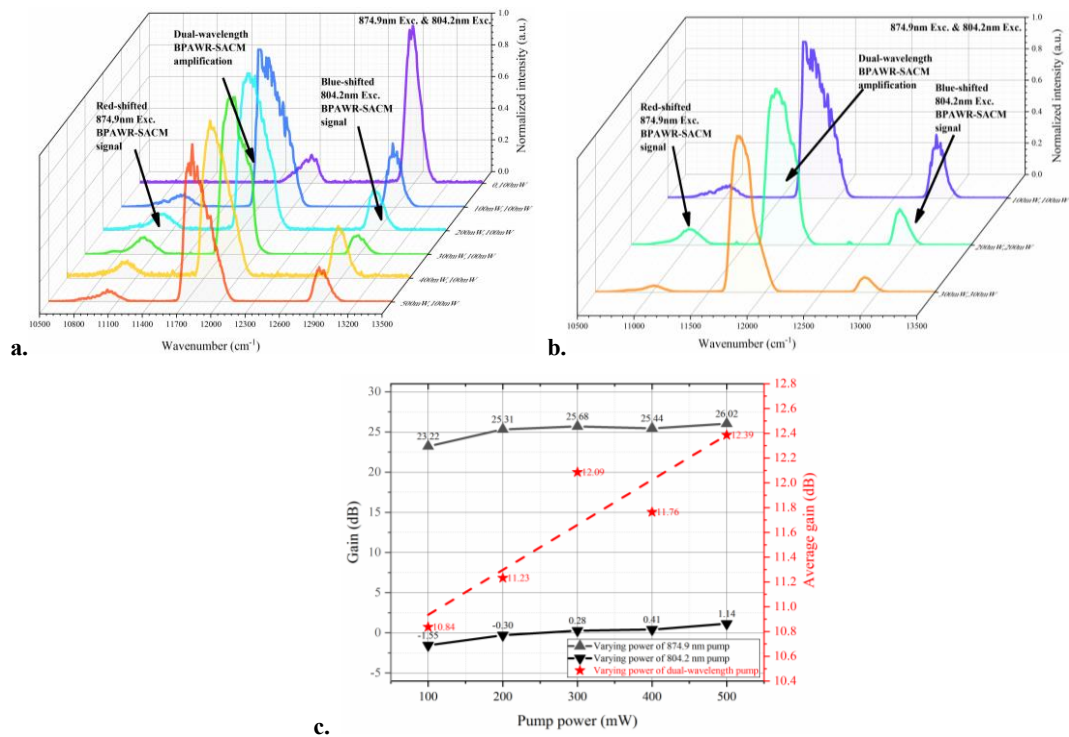

**Figure 4 | Signal amplification by the dual-wavelength pumping. a,** Normalized emission spectra obtained by maintaining the laser's power at 804.2 nm constant and varying the laser's power at 874.9 nm alone. **b,** Normalized emission spectra obtained by keeping the dual-wavelength pump power at the same level. **c,** The signal gain versus pump power. Gain was calculated from the amplification produced by keeping one laser power constant while increasing another laser power alone. Average gain was calculated from the amplification produced by synchronously increasing the dual-wavelength pump power.

These findings demonstrate that the significant phonon broadening in the HEGS plays an important role in generating the BPAWR process, which differs much from conventional phonon-assisted radiation due to the subsequent SACM process. Unlike conventional fluorescence, Raman scattering, or frequency multiplication process, this process shows spatial coherence at a low excitation threshold. The dynamic BPAWR-SACM process can be tuned by varying the excitation wavelengths, sample size, and doping concentration. The excitation wavelength influences the conversion efficiency of the BPAWR-SACM process. Signal amplification was demonstrated as one application of the BPAWR-SACM process. Multiple pump lights with different frequencies can amplify signal light, and therefore the frequency requirement for pump light will be much lower. Beyond this initial illustration, this work inspires us to design the BPAWR-SACM process in materials like semiconductors, carbon quantum dots, metal clusters, or materials with specific micro-nano scale structures, including multilayer coating systems nanostructures similar to moth-eyes, etc. [42]. Structure absorption affects the generation of the SACM, and BPAWR-SACM can achieve the modulation of the frequency, polarization or spin angular momentum with different conversion efficiencies. Moreover, BPAWR-SACM has a certain similarity with evanescent waves [43,44]. Such a result also attracts our interest in the BPAWR-SACM process generated at some high-entropy medium interfaces.

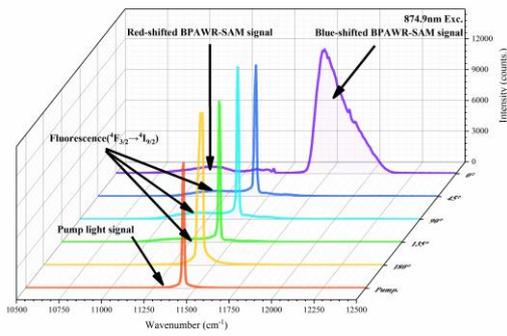 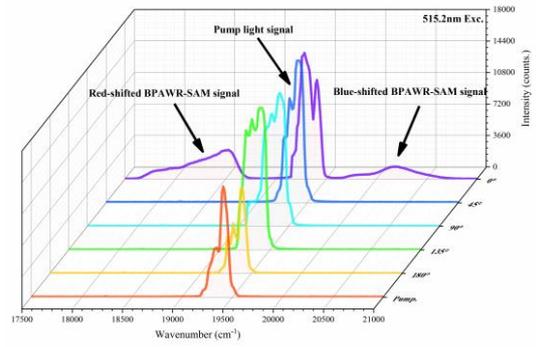

**Extended data Fig. 1 | Emission spectra at different measurement angles relative to the forward direction of the 874.9 nm excitation (a) and 515.2 nm excitation (b).** The corresponding excited energy levels are marked in the figure.

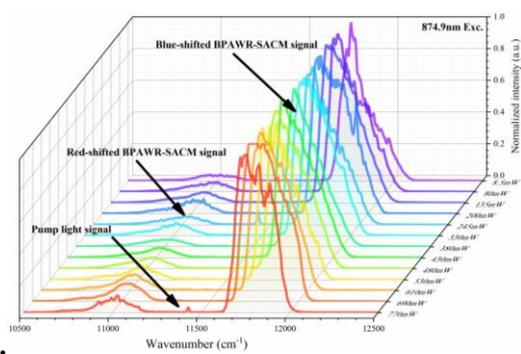 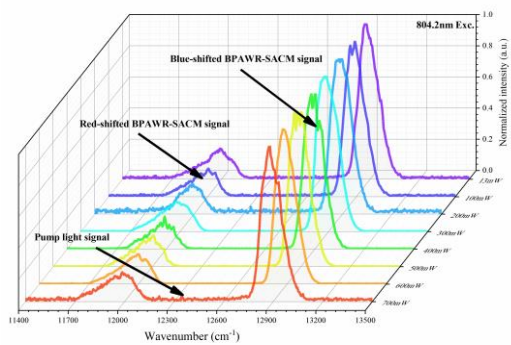

**Extended data Fig. 2 | Emission spectra obtained at different excitation powers of 874.9 nm laser (a) and 804.2 nm laser (b).** The corresponding BPAWR-SACM emission peak signal can be obtained with the laser power of 8.5 mW and 13 mW, respectively.

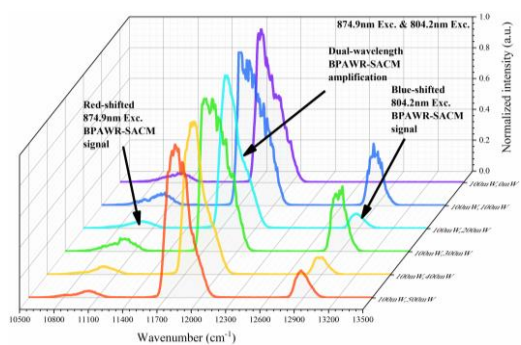

**Extended data Fig. 3 | Signal amplification by the dual-wavelength pumping.** Normalized emission spectra obtained by maintaining the laser's power at 874.9 nm constant and changing the power at 804.2 nm alone.

# Supplementary information

## 1. Supplementary Methods

### 1.1 Preparation of Nd$^{3+}$ doped HEGS

The recipe of HEGS includes 23.76 g zinc dihydrogen phosphate, 16 g phosphorus pentoxide, 13.92 g potassium sulfate, variable amounts of neodymium chloride (1.45 g, 2.90 g, 7.20 g, 8.71 g), 1.92 g lithium hydroxide, 0.9 g calcium chloride, 8.32 g tetraethyl silicate. The mass fractions of neodymium chloride with different amounts are 2.19 wt.%, 4.28 wt.%, 10 wt.%, and 11.84 wt.%, respectively. The mixture was melted at 700 °C to a flowing state and then poured into a mold. The corresponding glass samples were obtained by annealing, which was implemented: Firstly, lower the sample temperature from 700 °C to 300 °C at a cooling rate of 20 °C/min. Holding for a time, it then lowers the temperature from 300 °C to 250 °C at a cooling rate of 10°C/min. When the sample's temperature is uniform, cool the sample to 200 °C at the rate of 10°C/min. Finally, the sample is cooled down to room temperature at a rate of 1°C/min.

### 1.2 Characterization techniques

X-ray diffraction (XRD) patterns of HEGS samples were characterized by a powder diffractometer (X'Pert3 Powder) at a scanning rate of 5.7°/min and a step size of 0.026° (Co K$_\alpha$ radiation, λ=0.179 nm). Differential scanning calorimetry (DSC, SDT Q600) was carried out by heating 50 mg of glass powder in a nitrogen atmosphere at a heating rate of 10 °C/min. Microstructure characterization was performed by a transmission electron microscopy (TEM, FEI Tecnai G2 F30) with an accelerating voltage of 300 kV. With a HEGS sample of thickness 3 mm, its mid-infrared to near-infrared absorption spectrum was obtained by a UV-VIS-NIR spectrometer (Agilent, Cary 5000). And the far-infrared to mid-infrared transmission spectrum was measured by an FT-IR spectrometer (PerkinElmer, Spectrum Two).

### 1.3 Experimental methods

*High-power emission spectrum measurement*
The emission spectrum was measured by a fiber optic spectrometer from one end of a 65 mm HEGS sample pumped by a continuous wave (CW) semiconductor laser at 519.7 nm with a power of 1000 mW from another end. All the emission spectra were measured in this way, if not otherwise specified.

*Spatial coherence measurement*
An 804.2 nm laser of power 400 mW was used to excite one HEGS sample with a length of 35 mm. The fiber optic spectrometer's fiber port was placed on a circle of radius 20 mm, centered on the midpoint of the sample rod. By defining the side where the laser was located as 180° and the opposite side as 0°, the corresponding emission spectra were measured at an angle of 0° to 180° with an interval of 45°. Similar experiments were conducted with a 515.2 nm laser of power 400 mW and an 874.9 nm laser of power 400 mW.

*Time-delay dynamics measurement*
We employed a CW 532 nm laser to pump Ti: Sapphire and generated mode-locked 802.0 nm infrared light for excitation. The mode-locked laser pulses had a pulse width of 15 ps, one single pulse energy of 3 nJ, a repetition frequency of 86 MHz, and average power of 30 mW. As shown in supplementary Fig. 1, the infrared light was split into two beams with a 45° half mirror. The transmission light, serving as a reference signal, was measured by an oscilloscope following photodetector A. The reflection light was converted by photodetector B and measured by the oscilloscope if the sample was absent. Such a signal was referred to as a zero-delay signal. Using one HEGS sample with a length of 35 mm, we first confirmed its BPAWR-SAM characteristics with a fiber optic spectrometer and then measured the BPAWR-SAM signal by the oscilloscope.

*Threshold or BPAWR-SACM signal power measurement*
A CW 874.9 nm diode laser with a power of 8.5 mW to 770 mW was incident into a 35 mm HEGS sample. When the spectrum characteristics of the BPAWR-SACM process were observed and the excitation light was completely absorbed, the optical fiber spectrometer was replaced with a laser power meter to measure the power corresponding to the BPAWR-SACM signal. Similar experiments were conducted with an 804.2 nm laser with a power of 13 mW to 700 mW.

*Emission spectrum measurement at a single wavelength excitation with varying wavelengths*
A 35 mm HEGS sample was excited by lasers with wavelengths varying from 441.6 nm to 874.9 nm with a constant power of 400 mW. The corresponding emission spectra were measured with a fiber optic spectrometer.

*Emission spectrum measurement at a single wavelength excitation with varying sample sizes*
Using a 519.7 nm semiconductor laser with a power of 1000 mW as excitation, we tested the HEGS samples with lengths ranging from 10 mm to 90 mm. The corresponding emission spectra were measured with a fiber optic spectrometer.

*Emission spectrum measurement at a single wavelength excitation with varying doping concentration*
We employed a 1000 mW diode laser at 519.7 nm to excite 65 mm HEGS samples doped with $Nd^{3+}$ ions. The doping concentration ranged from 2.19 wt.% to 11.84 wt.%. The corresponding emission spectra were measured with a fiber optic spectrometer.

*Emission spectrum and signal power measurement at dual-wavelength excitation with varying excitation power*
A half mirror, CW diode lasers of 874.9 nm and 804.2 nm were employed to pump a 35 mm HEGS sample. The transmitted combined beam was used to generate the BPAWR-SACM process, and the emission spectra were measured with a fiber optic spectrometer. Then a laser power meter replacing the fiber optic spectrometer to measure the power of the BPAWR-SACM process. The incident light intensity was monitored in real-time with a laser power meter that received the signal from the reflected combined beam. The corresponding emission spectra and signal power were analyzed in three cases. Firstly, maintaining the pump power of 874.9 nm laser at 100mW, we increased the pump power of 804.2 nm laser from 0 mW to 500 mW. Then, the pump power of 804.2 nm laser was kept at 100mW, the pump power of 874.9 nm laser was adjusted from 0 mW to 500 mW. Lastly, the pump powers of both the 874.9 nm laser and the 804.2 nm laser were adjusted simultaneously.

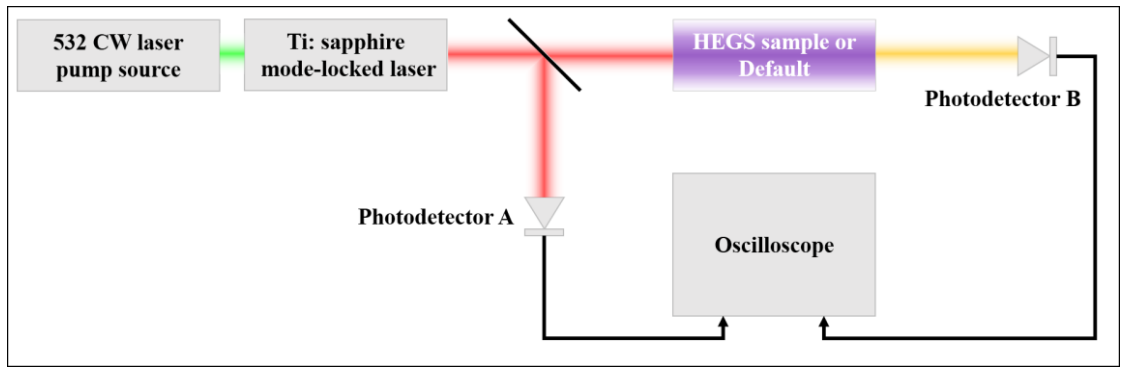

**Supplementary Fig. 1 | Optical configuration of time-delay dynamics measurement.** We employed a CW 532 nm laser to pump Ti: Sapphire and generated mode-locked 802.0 nm infrared light for excitation. The mode-locked laser pulses had a pulse width of 15 ps, one single pulse energy of 3 nJ, a repetition frequency of 86 MHz, and average power of 30 mW

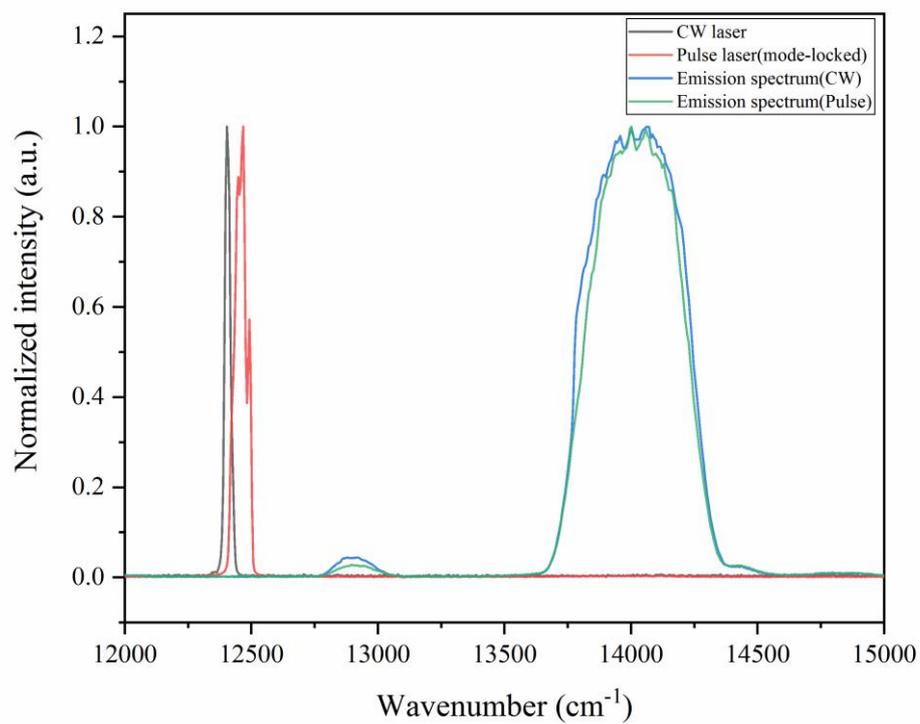

**Supplementary Fig. 2 | BPAWR-SAM spectra obtained during the time-delay measurements.** The emission wavelength of the CW laser before the mode-locked is 806.2nm, and the wavelength is shifted to 802.0 nm after the mode-locked. The relative peak positions of the emission peaks at the two wavelengths did not change significantly, and the emission peaks corresponding to BPAWR-SACM appeared at 12912.0 cm$^{-1}$ and 14002.3 cm$^{-1}$.

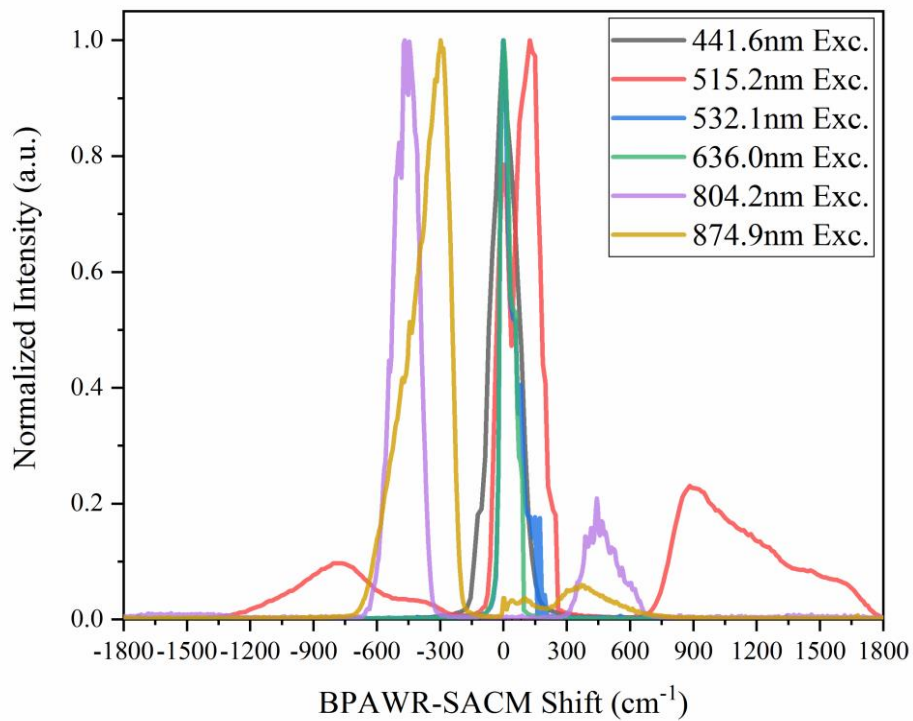

**Supplementary Fig. 3 | Normalized BPAWR-SACM shift spectra at different excitation wavelengths.** Only when the wavelength of the excitation light corresponds to the energy difference required for the energy level transition of $Nd^{3+}$ ions, a significant BPAWR-SACM signal can be generated. The red-shifted and blue-shifted signals are asymmetric, and the peak position shifts of the corresponding emission spectra at different excitation wavelengths are not consistent.